\documentstyle[12pt,eqsecnum,aps,fleqn,prb]{revtex}

\begin{document}

\title{ Asymptotic behaviours of the heat kernel in
covariant perturbation theory}
\vspace{10mm}
\author{\\
  A. O. Barvinsky\hbox{$^{1,\rm a)}$},
  Yu. V. Gusev\hbox{$^{1,\rm b)}$},
  G. A. Vilkovisky\hbox{$^{2}$} \\and\\
  V. V. Zhytnikov\hbox{$^{1,\rm c)}$}
 }

\maketitle
%\vspace{5mm}
\begin{center}
$^{1}$ {Nuclear Safety Institute, Bolshaya Tulskaya 52, Moscow
113191,
Russia}\\$^{2}${Lebedev Physics Institute, Leninsky Prospect 53,
Moscow 117924,
Russia}
\end{center}
%\bigskip
%\bigskip
\vspace{10mm}

\begin{abstract}
The trace of the heat kernel is expanded in a basis of nonlocal
curvature
invariants of $n$th order. The coefficients of this expansion (the
nonlocal
form factors) are calculated to third order in the curvature
inclusive. The
early-time and late-time asymptotic behaviours of the trace of the
heat kernel
are presented with this accuracy. The late-time behaviour gives the
criterion
of analyticity of the effective action in quantum field theory. The
latter
point is exemplified by deriving the effective action in two
dimensions.
\end{abstract}

\vspace{20mm}
PACS numbers: 04.60.+n, 11.15.-q, 02.40.+m\\

\thispagestyle{empty}
\footnoterule
\begin{itemize}
\item[$^{\rm a)}$]On leave at the Department
of Physics, University of Alberta, Edmonton T6G 2J1, Canada
\item[$^{\rm b)}$]
On leave at the Department of Physics,
University of Manitoba,
Winnipeg R3T 2N2, Canada
\item[$^{\rm c)}$]
On leave at the Department of Physics of
National Central University,
Chung-li, Taiwan 320
\end{itemize}

\pagebreak

\section{Introduction}

\indent
Heat kernel is a universal tool in theoretical and mathematical
physics. One can point out, in particular, its applications to
quantum theory
of gauge fields, quantum gravity \cite{Fock,1,2,5',18,3,4,5}, theory
of strings
\cite{6} and mathematical theory of differential operators on
nontrivial
manifolds \cite{7,8,9,10,11,21,12,24,Ram}. The significance of this
object
follows from the fact that the vast scope of problems boils down to
the
analysis of the
operator quantity
	\begin{equation}
	K(s)=\exp sH               \label{eqn:1.1}
	\end{equation}
which is associated with some differential operator $H$ and has a
kernel
	\begin{eqnarray}
	K(s|x,y)&\equiv&\exp sH\,\delta(x,y)   \label{eqn:1.2}
	\end{eqnarray}
solving the heat equation
	\begin{eqnarray}
	\frac{\partial}{\partial s}\,
	K(s|x,y)&=&H\,K(s|x,y),\,\,\,
	K(s|x,y)\,|_{s=0}=\delta\,(x,y).        \label{eqn:1.3}
	\end{eqnarray}
Here we shall focus on quantum field theory, in which case $H$
coincides with
the Hessian of the classical action (times a local matrix \cite{3}).
In quantum
field theory, the heat kernel (1.1) governs the semiclassical loop
expansion to
all orders, and in the covariant diagrammatic technique
\cite{1,2,3,4,5,13,14,15,16} it becomes indispensable. In particular,
it
generates the main ingredient of this expansion -- the propagator of
the
theory,
	\begin{equation}
	\frac{1}{H}=-\int_{0}^{\infty}ds\,K(s), \label{eqn:1.4}
	\end{equation}
and at one-loop order leads to the effective action in terms of the
functional
trace of the heat kernel ${\rm Tr} K(s)$
	\begin{eqnarray}
	&&\!\!\!\!-W=\frac12\int^\infty_0
	\frac{ds}s\,{\rm
	Tr}\,K(s),                       \label{eqn:1.5}\\
	&&{\rm Tr}\,K(s)\equiv\int
	d^{2\omega}x\;
	{\rm tr}\,K(s|x,y)\;\Big|_{y=x}.   \label{eqn:1.6}
	\end{eqnarray}
Here ${\rm tr}$, as distinct from ${\rm Tr}$ in (1.5), denotes the
matrix trace
with respect to the discrete indices of $K(s|x,y)$ (which
arise for any matrix-valued operator $H$, corresponding to the fields
of
nonzero spin) and $2\omega$ is the spacetime dimension.

Apart from special backgrounds (see, e.g.,\cite{12}), the heat kernel
cannot be
calculated exactly, and for the effective action in quantum field
theory, one
needs it as a {\it functional} of background fields \cite{18,5}. For
an
approximate calculation, two regular schemes \cite{F} have thus far
been used:
the technique of asymptotic expansion at early time
\cite{2,7,8,9,10,17}
$(s\rightarrow 0)$ and covariant perturbation theory
\cite{13,14,15,16}.

As seen from (1.5), the asymptotic expansion of (1.6) in
$s\rightarrow 0$
allows one to single out the ultraviolet divergences generated by
the lower integration limit in $s$. One can also obtain the local
effects of
vacuum
polarization by massive quantum fields \cite{Frolov-Zeln}
corresponding to the
asymptotic
expansion of the effective action in inverse powers of large mass
parameter
$m^2\rightarrow\infty$. In massive theories, the integral
(1.5) acquires the multiplier $e^{-m^2 s}$ exponentially suppressing
large scales of $s$ and generating the $1/m^2-$expansion.
The coefficients of this expansion are the spacetime integrals of
local
invariants
$a_{n}(x,x),\;n=0,1,2,...$, of growing power in curvature and its
derivatives,
called DeWitt or HAMIDEW coefficients. The local (Schwinger-DeWitt)
expansion
is, therefore, an asymptotic approximation of {\it small and slowly
varying}
background fields.
The term HAMIDEW, which means Hadamard-Minakshisundaram-DeWitt, has
been proposed by G.Gibbons \cite{Gibbons} to praise the joint efforts
of
mathematicians and physicists in the pioneering studies of the
early-time
expansion of the heat
kernel \cite{7,8,2}. These studies contained the explicit calculation
of
$a_{n}(x,x)$ for $n=0,1$ and $2$. The coefficient $a_{3}(x,x)$ has
been worked
out by P.B.Gilkey
\cite{10},
while the highest-order coefficient available now for a generic
theory,
$a_{4}(x,x)$, was obtained by I.G.Avramidy \cite{17} (see also
Ref.\onlinecite{Amsterdamski}).

This expansion, very efficient for obtaining covariant
renormalizations and
anomalies, becomes, however, unreliable for large and/or rapidly
varying
fields and completely fails in massless theories, because, in the
absence of a damping factor $e^{-m^2 s}$, the early-time expansion
of ${\rm Tr}\,K(s)$
in (1.5) is non-integrable at the upper limit $s\rightarrow\infty$.
The calculational technique which solves this problem in the case of
rapidly
varying fields is covariant perturbation
theory \cite{13,14,15,16}. It corresponds to a partial summation of
the
Schwinger-DeWitt series by summing all terms with a given power of
the curvature and any number of derivatives \cite{f2}. This is still
an
expansion of ${\rm Tr}\,K(s)$
in powers of the curvature but the coefficients of this expansion are
nonlocal.
In contrast to the Schwinger-DeWitt expansion, in this
technique the convergence of the integral (1.5) at $s\rightarrow
\infty$ for
massless theories is controlled by the late-time behaviour of these
nonlocal
coefficients, which altogether comprise the late-time behaviour of
the heat
kernel.

Covariant perturbation theory is already capable of reproducing the
effects of
nonlocal vacuum polarization and particle creation by rapidly varying
fields.
It should, therefore, contain the Hawking radiation effect
\cite{Hawking} and
its backreaction on the metric in the gravitational collapse problem
\cite{Fr-Vil,Vil-Tex,CQG}. This was in fact the original motivation
for
studying this theory. This motivation has recently been strengthened
in the
work \cite{Armen} where it is shown that loop expansion of field
theory can be
trusted in the spacetime domain near null infinity where the massless
vacuum
particles are radiated.

Covariant perturbation theory was proposed in \cite{13} and then
applied for
the calculation of the heat kernel and effective action to second
order in the
curvature \cite{14}. However, as pointed out in papers
\cite{CQG,5,Armen}, one
has to go as far as third order in the curvature (in the action) to
obtain the
component of radiation that remains stable after the black hole is
formed. The
third-order calculation has recently been completed \cite{16} for
both the heat
kernel and one-loop effective action. The basis of third-order
curvature
invariants was built \cite{16,JMP}, and all nonlocal coefficients of
these
invariants both in the heat kernel and effective action were
calculated as
functions of three commuting operator arguments. Several integral
representations for these functions were obtained. The results for
the
effective action checked by deriving the trace anomaly in two and
four
dimensions. Here we focuse on the results for the heat kernel.

The structure of nonlocal coefficients in the heat kernel is
discussed below
but the full results for these coefficients \cite{16} are usable only
in the
format of the computer algebra program {\it Mathematica}
\cite{files}.
Therefore, they will not be presented here. Below we present only the
asymptotic behaviours of the heat kernel at early and late times,
which are
most important, in view of the discussion above, for the theory of
massless
quantum fields, and for the spectral analysis on Riemann manifolds.

For a technical discussion of covariant perturbation theory and
relevant
physical problems we refer the reader to papers \cite{13,14,15} and
the recent
paper \cite{16,JMP} where this theory along with some of its
applications is
reviewed. Although we consider only the trace of the heat kernel, the
knowledge
of the functional trace (1.6) is sufficient, owing to the variational
method in
\cite{20}, for obtaining also the coincidence limit of the heat
kernel
${\rm tr}\,K(s|x,y)\;\Big|_{y=x}$ and coincidence limits of its
derivatives
with respect to one of the spacetime arguments. Finally, by using the
method in
\cite{21,22}, covariant perturbation theory can be extended to the
calculation
of heat kernel with separated points -- the object very important for
multi-loop diagrams \cite{4} and for field theory at finite
temperature
\cite{23a}. To second order in the curvature, this object was
calculated in
\cite{23b}.

Like the Schwinger-DeWitt technique, covariant perturbation theory
fails in the
case of large fields. Apart from some special cases \cite{Jack-P},
the problem
of large fields remains unsolved but one may think of an
approximation scheme
complimentary to covariant perturbation theory, where one starts with
large and
slowly varying fields. The lowest-order approximation of such a
scheme would be
the case of covariantly constant background fields. This
approximation has
recently been considered (for the vanishing Riemann curvature,
however) in
\cite{25} with a result generalizing the old Schwinger's result
\cite{1} to the
non-abelian vector fields. In a rather non-trivial way, the result in
\cite{25}
can also be extended to the case of non-vanishing, covariantly
constant,
Riemann curvature \cite{Avramidi}.

The plan of the paper is as follows. In Sect.II we introduce the
notation, and
review the general setting of covariant perturbation theory for the
heat kernel
in the case of a generic second-order Hamiltonian. Next we comment on
the full
results obtainede for the trace of the heat kernel to third order in
the
curvature. Sect.III presents the early-time asymptotic behaviour of
the trace
of the heat kernel as obtained from these results. We carry out a
comparison
with the Schwinger-DeWitt series and, as a by-product, obtain a
workable
expression for the cubic terms of the HAMIDEW coefficient $a_4$.
Finally,
Sect.IV contains the main result of the present paper: the late-time
asymptotic
behaviour of the trace of the heat kernel to third order in the
curvature. As
mentioned above, the capability of covariant perturbation theory of
producing
this behaviour is the principal advantage of the method.

\section{Covariant perturbation theory for the
trace of the heat kernel}
\indent
The subject of calculation in covariant perturbation theory of
\cite{13,14,15,16} is the heat kernel (1.1) where $H$ is the generic
second-order operator
	\begin{equation}
	H=g^{\mu\nu}\nabla_\mu\nabla_\nu\hat{1}+
	(\hat{P}-\frac16R\hat{1}),
	\hspace{7mm}
	g^{\mu\nu}\nabla_\mu\nabla_\nu
	\equiv\Box                             \label{eqn:2.1}
	\end{equation}
acting in a linear space of fields $\varphi^A(x)$.
Here $A$ stands for any set of discrete
indices, and the hat indicates that the quantity is a matrix
acting on a vector $\varphi^A$:
	$\hat{1}=
	\delta^A{}_B,$ $\hat{P}=P^A{}_B$,
etc. We have
	${\rm tr}\hat{1}=
	\delta^A{}_A$, ${\rm tr}\hat{P}=P^A{}_A$,
etc. In (2.1), $g_{\mu\nu}$ is a positive-definite metric
characterized by its Riemann curvature \cite{f4}
$R^{\alpha\beta\mu\nu}$,
$\nabla_\mu$ is a covariant derivative (with respect to an
arbitrary connection) characterized by its commutator
curvature
	\begin{equation}
	(\nabla_\mu\nabla_\nu-\nabla_\nu\nabla_\mu)
	\varphi^A =
	{\cal R}^A{}_{B\mu\nu}\varphi^B,\hspace{7mm}
	{\cal R}^A{}_{B\mu\nu}
	\equiv\hat{\cal R}_{\mu\nu},           \label{eqn:2.4}
	\end{equation}
and $\hat{P}$ is an arbitrary matrix. The redefinition of the
potential in (2.1) by inclusion of the term in the Ricci
scalar $R$ is a matter of convenience, while the absence of the term
linear in
$\nabla_\mu$ can be always achieved by a redefinition of the
connection
entering the covariant derivative. The only important limitation in
the
operator (2.1) is the structure of its second-order
derivatives which
form a covariant Laplacian $\Box$. There exist, however, efficient
methods
\cite{3} by which a more general problem can be reduced to the case
of (2.1)
(see also \cite{24}).

The present paper, like the preceding ones \cite{13,14,15,16}, deals
only
with the version of covariant perturbation theory appropriate
for noncompact asymptotically flat and empty manifolds. The interest
in this setting follows from
the fact
that it results in the Euclidean effective action which, for certain
quantum
states, is sufficient for obtaining the scattering amplitudes and
expectation-value equations of Lorentzian field theory \cite{13}.
Under the assumed conditions, the covariant Laplacian $\Box$,
has a
unique Green function corresponding to zero boundary conditions at
infinity.
The masslessness of the operator (2.1) means that, like the
Riemann and commutator curvatures, the potential $\hat{P}$
falls off at infinity. For the
precise conditions of this fall off see \cite{14}.

For the set of the field strengths (curvatures)
	\begin{equation} R^{\alpha\beta\mu\nu},
	\hspace{7mm}
	\hat{\cal R}_{\mu\nu},
	\hspace{7mm} \hat{P}                \label{eqn:2.5}
	\end{equation}
characterizing the background we use the collective
notation $\Re$. The calculations in covariant perturbation
theory are carried out with accuracy ${\rm O}[\Re^N]$, i.e.
up to terms of $N$th and higher power in the curvatures
(\ref{eqn:2.5}).
It is worth noting that, since the calculations are covariant,
any term containing the metric is in fact of infinite power in the
curvature, and ${\rm O}[\Re^N]$ means terms containing $N$ or
more curvatures {\em explicitly}.

The trace of the heat kernel is an invariant functional of background
fields
which belongs to the class of invariants considered in \cite{JMP}. It
is,
therefore, expandable in the basis of nonlocal invariants of $N$th
order built
in \cite{JMP}. To third order, the expansion is of the form
\begin{eqnarray}
{\rm Tr}\,K(s)\!\!\!&=&
\frac1{(4\pi s)^\omega}\int
dx\, g^{1/2}\, \,{\rm tr}\,
\Big\{\,\hat{1}+s\hat{P}
%\nonumber\\&&
+s^2\sum^{5}_{i=1}f_{i}(-s\Box_2)\,
\Re_1\Re_2({i})\nonumber\\&&
+s^3\sum^{11}_{i=1}F_{i}(-s\Box_1,-s\Box_2,-s\Box_3)\,
\Re_1\Re_2\Re_3({i})\nonumber\\&&
+s^4\sum^{25}_{i=12}F_{i}(-s\Box_1,-s\Box_2,-s\Box_3)\,
\Re_1\Re_2\Re_3({i})\nonumber\\&&
+s^5\sum^{28}_{i=26}F_{i}(-s\Box_1,-s\Box_2,-s\Box_3)\,
\Re_1\Re_2\Re_3({i})\nonumber\\&&
+s^6\,\phantom{\sum^{28}_{i=26}}\!\!\!\!\!\!\!\!\!
F_{29}(-s\Box_1,-s\Box_2,-s\Box_3)\,
\Re_1\Re_2\Re_3({29})\,
+{\rm O}[\,\Re^4\,]\,\Big\}.             \label{eqn:2.6}
\end{eqnarray}
The full list of quadratic $\Re_1\Re_2({i}),\;i=1\;{\rm to}\;5,$ and
cubic
$\Re_1\Re_2\Re_3({i}),\;i=1\;{\rm to}\;29,$ curvature invariants
here, as well
as the conventions concerning the action of box operator arguments
$(\Box_1,\Box_2,\Box_3)$ on the curvatures $(\Re_1,\Re_2,\Re_3)$
labelled by
the corresponding numbers, are presented and discussed in much detail
in
\cite{JMP}. The form factors of this expansion $f_{i}(\xi)$, $i=1$ to
$5$,  and
$F_{i}(\xi_1,\xi_2,\xi_3)$, $i=1$ to $29$, all express through the
basic
second-order and third-order form factors
	\begin{eqnarray}
	&&\!\!\!\!f(\xi)=\int_{\alpha\geq0}\!d^2\alpha\,
	\delta(1-\alpha_1-\alpha_2)
	\exp(-\alpha_1\alpha_2\xi)
	=\int^1_0\!d\alpha\,
	{\rm e}^{-\alpha(1-\alpha)\xi},  \label{eqn:2.7}
	\\
	&&\!\!\!\!F(\xi_1,\xi_2,\xi_3)=
	\int_{\alpha\geq0}d^3\alpha\,
	\delta(1-\alpha_1-\alpha_2-\alpha_3)
	\exp(-\alpha_1\alpha_2\xi_3
	-\alpha_2\alpha_3\xi_1
	-\alpha_1\alpha_3\xi_2).                   \label{eqn:2.8}
	\end{eqnarray}
The structure of these expressions is as follows:
	\begin{eqnarray}
	&&\!\!\!\!f_{i}(\xi)=r_i(\xi)\,f(\xi)+v_i(\xi),
\label{eqn:2.9}\\
	&&\!\!\!\!F_{i}(\xi_1,\xi_2,\xi_3)=
	R_{i}(\xi_1,\xi_2,\xi_3)\,F(\xi_1,\xi_2,\xi_3)+
	\sum_{n=1}^{3}\,U_{i}^{n}(\xi_1,\xi_2,\xi_3)\,
	f(\xi_n)+V_{i}(\xi_1,\xi_2,\xi_3)       \label{eqn:2.10}
	\end{eqnarray}
where $r_i(\xi)$, $v_i(\xi)$, $R_{i}(\xi_1,\xi_2,\xi_3)$,
$U_{i}^{n}(\xi_1,\xi_2,\xi_3)$ and $V_{i}(\xi_1,\xi_2,\xi_3)$
are certain rational functions of their arguments. The functions
$r_i$ and
$v_i$ have in the denominator the powers of $\xi$, while the
denominators of
$R_{i}$, $U_{i}^{n}$ and $V_{i}$ are formed by the powers of the
universal
quantity $\xi_1\xi_2\xi_3\Delta$,
$\Delta\equiv\xi_1^2+\xi_2^2+\xi_3^2-2\xi_1\xi_2-2\xi_2\xi_3-2
\xi_3\xi_1$, and
also contain the factors $(\xi_j-\xi_k)$, $j\not=k$, (in the
denominators of
$U_{i}^{n}$). Despite this fact, all the form factors are analytic in
their
arguments at $\xi_j=0$, $\Delta=0$ and $\xi_j=\xi_k$, and the
mechanism of
maintaining this analyticity is based on linear differential
equations which
the functions (\ref{eqn:2.7}) -- (\ref{eqn:2.8}) satisfy \cite{16}.
Here we
present some of these equations in the form which will be used in
Sect.IV for
the derivation of the effective action for massless conformal field
in two
dimensions:
\begin{eqnarray}
&&\frac{d}{d\xi}f(\xi)=-\frac14 f(\xi)-\frac12
\frac{f(\xi)-1}{\xi},                \label{eqn:2.11}\\
&&\frac{d^2}{d\xi^2}f(\xi)=\frac1{16}f(\xi)
+\frac14\frac{f(\xi)-1}{\xi}
+\frac34\frac{f(\xi)-1+\frac16\xi}{\xi^2},\label{eqn:2.12}\\
&&s\frac{\partial}{\partial s}F(-s\Box_1,-s\Box_2,-s\Box_3)=
-\left(s\frac{\Box_1\Box_2\Box_3}{D}+1
\right)F(-s\Box_1,-s\Box_2,-s\Box_3)
\nonumber\\&&\qquad\qquad
-\frac{\Box_1(\Box_3+\Box_2-\Box_1)}{2D}f(-s\Box_1)
-\frac{\Box_2(\Box_3+\Box_1-\Box_2)}{2D}f(-s\Box_2)
\nonumber\\&&\qquad\qquad
-\frac{\Box_3(\Box_1+\Box_2-\Box_3)}{2D}f(-s\Box_3),\label{eqn:2.13}
\\&&
D\equiv{\Box_1}^2+{\Box_2}^2+{\Box_3}^2-2\Box_1\Box_2
-2\Box_1\Box_3-2\Box_2\Box_3.                        \label{eqn:2.14}
\end{eqnarray}

The explicit results for the second-order form factors are given in
\cite{14}
(see also \cite{17}). The explicit results for the third-order
form factors
are given in \cite{16} and take pages. They are, however, manageable
in the
format of the computer algebra program {\it Mathematica}. This
program was used
for a number of derivations in \cite{16}, and for obtaining the
asymptotic
behaviours presented below.

\section{The early-time behaviour of the trace
of the heat kernel and the Schwinger-DeWitt
expansion}
%\indent
The early-time behaviour of the heat kernel follows from the
tables of
paper \cite{16} and  the behaviour of the basic form factors
(\ref{eqn:2.7})
and (\ref{eqn:2.8})
	\begin{equation}
f(-s\Box)=1 + \frac16 s\Box +\frac{1}{60}s^2 \Box^2
+{\rm O}\left(s^3\right),
\ \ \ \  s\rightarrow 0                    \label{eqn:3.1}
	\end{equation}
	\begin{equation}
F(-s\Box_1,-s\Box_2,-s\Box_3) =
\frac12 +\frac{1}{24}s({\Box_1}
+{\Box_2}+\Box_3) +{\rm O}\left(s^2\right),
\ \  s\rightarrow 0.                      \label{eqn:3.2}
	\end{equation}
It is striking that, in the resulting early-time expansion, the
third-order
form factors are nonlocal and, for some of them, the expansion
starts with a negative power of $s$. By making a comparison with the
table of tensor structures of eq.(\ref{eqn:2.6}) presented in
\cite{16,JMP},
one can see that such a
behaviour is inherent only in the gravitational form factors, and,
moreover, the nonlocal operators $1/\Box$ in the asymptotic
expressions above act only on the gravitational curvatures. As
discussed in \cite{14,JMP}, these features will persist at all higher
orders in $\Re$, and the cause is that the basis set of curvatures
for
the heat kernel does not contain the Riemann tensor which gets
excluded via the
Bianchi identitites in terms of the Ricci tensor. Below we show that
restoring
the Riemann tensor restores the locality of the early-time expansion.

The early-time expansion for ${\rm Tr} K(s)$ is of the form \cite{2}
	\begin{equation}
	{\rm Tr} K(s)=\frac1{(4\pi s)^\omega}
	\sum_{n=0}^{\infty} s^n \int\! dx\, g^{1/2}\,
	{\rm tr}\ \hat{a}_n (x,x)          \label{eqn:3.3}
	\end{equation}
where $\hat{a}_n (x,x)$ are the DeWitt coefficients with coincident
arguments. All $\hat{a}_n (x,x)$ are {\em local} functions of the
background fields entering the operator (2.1). There exist
independent methods for obtaining these coefficients, and, for
$n=0,1,2,3,4$, the $\hat{a}_n (x,x)$ have been calculated explicitly
\cite{2,5',7,8,9,10,3,17,Amsterdamski}. A comparison with these known
expressions, carried out below, provides a powerful check of the
results of
covariant perturbation theory.

By using the behaviours (\ref{eqn:3.1}) -- (\ref{eqn:3.2}) and the
tables of
form factors of Ref.\onlinecite{16}, one arrives at
eq.(\ref{eqn:3.3}) with the following results for the (integrated)
DeWitt
coefficients $a_0$ to $a_4$:
\mathindent = 0pt
\arraycolsep=0pt
\begin{eqnarray}&&
\int\! dx\, g^{1/2}\,
{\rm tr}\, \hat{a}_0 (x,x)=\int\! dx\,
g^{1/2}\,  {\rm tr}\, \hat{1},              \label{eqn:3.4}
\end{eqnarray}
\begin{eqnarray}&&
\int\! dx\, g^{1/2}\,  {\rm tr}\,
\hat{a}_1 (x,x)=\int\! dx\,
g^{1/2}\,  {\rm tr}\, \hat{P},              \label{eqn:3.5}
\end{eqnarray}
\begin{eqnarray}&&
\int\! dx\, g^{1/2}\,  {\rm tr}\,\hat{a_2}(x,x)=
\int\! dx\, g^{1/2}\,  {\rm tr}\left\{ \frac12\hat{P}_1\hat{P}_2+
\frac1{12}\hat{\cal R}_{1\mu\nu}\hat{\cal R}_2^{\mu\nu}
+\frac1{60}R_{1\,\mu\nu} R_2^{\mu\nu}\hat{1}
-\frac1{180}R_1 R_2\hat{1}
\right.\nonumber\\&&
+{{\Box_3}\over {360 \Box_1 {\Box_2}}}
R_1 R_2 R_3\hat{1}
+\frac1{90}\Big(-{{2}\over {\Box_3}}
 + {{\Box_3}\over { \Box_1 \Box_2}}
\Big)
R_{1\,\alpha}^\mu R_{2\,\beta}^{\alpha} R_{3\,\mu}^\beta\hat{1}
+\frac1{180}\Big({2\over{\Box_2}}-{{\Box_3}\over{\Box_1\Box_2}}
\Big)
R_1^{\mu\nu}R_{2\,\mu\nu}R_3\hat{1}
\nonumber\\&&
+\frac1{180}\Big(-{{2}\over{\Box_1\Box_2}}-{3\over{\Box_2\Box_3}}
\Big)
R_1^{\alpha\beta}\nabla_\alpha R_2 \nabla_\beta R_3\hat{1}
+{1\over {45 \Box_1 \Box_2}}
\nabla^\mu R_1^{\nu\alpha}\nabla_\nu R_{2\,\mu\alpha}R_3\hat{1}
\nonumber\\&&
+{1\over {45 \Box_2 \Box_3}}
R_1^{\mu\nu}\nabla_\mu R_2^{\alpha\beta}\nabla_\nu
R_{3\,\alpha\beta}\hat{1}
+\frac1{45}\Big({2\over {\Box_1 \Box_2}}
- {1\over {\Box_2 \Box_3}}\Big)
R_1^{\mu\nu}\nabla_\alpha R_{2\,\beta\mu}
\nabla^\beta R_{3\,\nu}^\alpha\hat{1}
\nonumber\\&&
-{{1}\over {45 \Box_1 \Box_2 \Box_3}}
\nabla_\alpha\nabla_\beta R_1^{\mu\nu}
\nabla_\mu\nabla_\nu R_2^{\alpha\beta} R_3\hat{1}
\left.
-{{2}\over {45 \Box_1 \Box_2 \Box_3}}
\nabla_\mu R_1^{\alpha\lambda} \nabla_\nu
R_{2\,\lambda}^\beta\nabla_\alpha\nabla_\beta
R_3^{\mu\nu}\hat{1} \right\}
+{\rm O}[\Re^4],
\label{eqn:3.6}
\end{eqnarray}
\begin{eqnarray}&&
\int\! dx\, g^{1/2}\,  {\rm tr}\,\hat{a_3}(x,x)=
\int\! dx\, g^{1/2}\, {\rm tr}
\left\{\frac{\Box_2}{12}\hat{P}_1\hat{P}_2
+\frac{\Box_2}{120}\hat{\cal R}_{1\mu\nu}
\hat{\cal R}_2^{\mu\nu}+\frac{\Box_2}{180}\hat{P}_1 R_2
\right.\nonumber\\&&
+\frac{\Box_2}{840}R_{1\,\mu\nu} R_2^{\mu\nu}\hat{1}
-\frac{\Box_2}{3780}R_1 R_2\hat{1}
+{1\over 6}
\hat{P}_1\hat{P}_2\hat{P}_3
-{1\over {45}}
\hat{\cal R}^{\ \mu}_{1\ \alpha}
\hat{\cal R}^{\ \alpha}_{2\ \beta}\hat{\cal R}^{\ \beta}_{3\ \mu}
+{1\over {12}}
\hat{\cal R}^{\mu\nu}_1\hat{\cal R}_{2\,\mu\nu}\hat{P}_3
\nonumber\\&&
+\frac1{180}\Big({1} + {2{\Box_1}\over {\Box_2}}
- {4{\Box_3}\over {\Box_2}} +
  {{{{\Box_3}^2}}\over {\Box_1 \Box_2}}\Big)
  R_1^{\mu\nu}R_{2\,\mu\nu}\hat{P}_3
+\Big({1\over {45}} + {{\Box_3}\over {18 \Box_1}}\Big)
R_1^{\alpha\beta}\hat{\cal R}_{2\,\alpha}^{\ \ \ \,\mu}
\hat{\cal R}_{3\,\beta\mu}
\nonumber\\&&
+\frac1{3360}\Big(-{28\over27} + {4{\Box_1}\over{3\Box_3}} +
  {{{{\Box_3}^2}}\over{\Box_1\Box_2}}\Big)
R_1 R_2 R_3\hat{1}
+\frac1{840}\Big(-{4\over9} - {2{\Box_1}\over {\Box_3}} +
  {{{{\Box_3}^2}}\over {\Box_1 \Box_2}}\Big)
R_{1\,\alpha}^\mu R_{2\,\beta}^{\alpha} R_{3\,\mu}^\beta\hat{1}
\nonumber\\&&
+\frac1{15120}\Big({1} - {{\Box_1}\over {\Box_2}}
+ {8{\Box_3}\over{\Box_2}} -
  {{7{{\Box_3}^2}}\over{2\Box_1 \Box_2}}\Big)
R_1^{\mu\nu}R_{2\,\mu\nu}R_3\hat{1}
+\frac1{45}\Big({2\over{\Box_2}} - {{\Box_3}\over{\Box_1
\Box_2}}\Big)
\nabla^\mu R_1^{\nu\alpha}\nabla_\nu R_{2\,\mu\alpha}\hat{P}_3
\nonumber\\&&
+{1\over {18 \Box_1}}
R_{1\,\alpha\beta}\nabla_\mu\hat{\cal R}_2^{\mu\alpha}
\nabla_\nu\hat{\cal R}_3^{\nu\beta}
-{{1}\over {36 \Box_1}}
R_1^{\alpha\beta}\nabla_\alpha\hat{\cal R}_2^{\mu\nu}
\nabla_\beta\hat{\cal R}_{3\,\mu\nu}
+{1\over {9 \Box_1}}
R_1^{\mu\nu}\nabla_\mu\nabla_\lambda
\hat{\cal R}_2^{\lambda\alpha}\hat{\cal R}_{3\,\alpha\nu}
\nonumber\\&&
+\frac1{2520}\Big({1\over{\Box_1}} - {2\over{\Box_2}}
- {4{\Box_1}\over{\Box_2 \Box_3}} -
  {2{\Box_3}\over {\Box_1 \Box_2}}\Big)
R_1^{\alpha\beta}\nabla_\alpha R_2 \nabla_\beta R_3\hat{1}
\nonumber\\&&
+\frac1{1080}\Big( {{\Box_3}\over {\Box_1 \Box_2}}-{{2}\over
{7\Box_2}}
\Big)
\nabla^\mu R_1^{\nu\alpha}\nabla_\nu R_{2\,\mu\alpha}R_3\hat{1}
+\frac1{504}\Big({{\Box_1}\over {\Box_2 \Box_3}}
-{{6}\over {5 \Box_2}}\Big)
R_1^{\mu\nu}\nabla_\mu R_2^{\alpha\beta}
\nabla_\nu R_{3\,\alpha\beta}\hat{1}
\nonumber\\&&
+\frac1{420}\Big({2{\Box_3}\over { \Box_1 \Box_2}}-{{4}\over
{3\Box_1}}
- {{\Box_1}\over { \Box_2 \Box_3}} \Big)
R_1^{\mu\nu}\nabla_\alpha R_{2\,\beta\mu}
\nabla^\beta R_{3\,\nu}^\alpha\hat{1}
+{1\over {45 \Box_1 \Box_2}}
\nabla_\alpha\nabla_\beta R_1^{\mu\nu}
\nabla_\mu\nabla_\nu R_2^{\alpha\beta}\hat{P}_3
\nonumber\\&&
+\frac1{756}\Big(-{{1}\over {\Box_1 \Box_2}}
- {3\over {\Box_2 \Box_3}}\Big)
\nabla_\alpha\nabla_\beta R_1^{\mu\nu}
\nabla_\mu\nabla_\nu R_2^{\alpha\beta} R_3\hat{1}
\nonumber\\&&
+\frac1{315}\Big(-{{1}\over {\Box_1 \Box_2}}
- {3\over {\Box_2 \Box_3}}\Big)
\nabla_\mu R_1^{\alpha\lambda}
\nabla_\nu R_{2\,\lambda}^\beta
\nabla_\alpha\nabla_\beta R_3^{\mu\nu}\hat{1}
\nonumber\\&&\left.
+{1\over {1890 \Box_1 \Box_2 \Box_3}}
 \nabla_\lambda\nabla_\sigma R_1^{\alpha\beta}
\nabla_\alpha\nabla_\beta R_2^{\mu\nu}
\nabla_\mu\nabla_\nu
R_3^{\lambda\sigma}\hat{1} \right\}
+{\rm O}[\Re^4],                                     \label{eqn:3.7}
\end{eqnarray}
\begin{eqnarray}&&
\int\! dx\, g^{1/2}\,  {\rm tr}\,\hat{a_4}(x,x)=
\int\! dx\, g^{1/2}\, {\rm tr}\left.
\{\frac{{\Box_2}^2}{120}\hat{P}_1\hat{P}_2
+\frac{{\Box_2}^2}{1260}\hat{P}_1 R_2
+\frac{{\Box_2}^2}{1680}\hat{\cal R}_{1\mu\nu}
\hat{\cal R}_2^{\mu\nu}
\right.
\nonumber\\&&
+\frac{{\Box_2}^2}{15120}R_{1\,\mu\nu} R_2^{\mu\nu}\hat{1}
+{{\Box_3}\over {24}}
\hat{P}_1\hat{P}_2\hat{P}_3
-{{\Box_3}\over {630}}
\hat{\cal R}^{\ \mu}_{1\ \alpha}
\hat{\cal R}^{\ \alpha}_{2\ \beta}\hat{\cal R}^{\ \beta}_{3\ \mu}
+\frac{\Box_1 + \Box_2 + 2\Box_3}{180}
\hat{\cal R}^{\mu\nu}_1\hat{\cal R}_{2\,\mu\nu}\hat{P}_3
\nonumber\\&&
+\frac{2\Box_1 - \Box_3}{15120}
R_1 R_2 \hat{P}_3
+\frac1{1680}\Big(\Box_1
+ {{{{\Box_1}^2}}\over {\Box_2}} +
  {2{\Box_3}\over {3}} + {{\Box_1 \Box_3}\over {\Box_2}} -
  {{{5{\Box_3}^2}}\over { \Box_2}}
+ {{{3{\Box_3}^3}}\over {2 \Box_1 \Box_2}}\Big)
R_1^{\mu\nu}R_{2\,\mu\nu}\hat{P}_3
\nonumber\\&&
+{{\Box_3}\over {720}}
\hat{P}_1\hat{P}_2 R_3
+\frac1{840}\Big(\Box_1 + 4\Box_3
+ {4{\Box_2 \Box_3}\over {\Box_1}} +
  {{4{{\Box_3}^2}}\over {\Box_1}}\Big)
R_1^{\alpha\beta}\hat{\cal R}_{2\,\alpha}^{\ \ \ \,\mu}
\hat{\cal R}_{3\,\beta\mu}
\nonumber\\&&
+\frac{13 \Box_1 - 2\Box_3}{30240}
R_1\hat{\cal R}^{\mu\nu}_2\hat{\cal R}_{3\,\mu\nu}
+\frac1{50400}\Big({{{2{\Box_1}^2}}\over {\Box_3}}
+ {{\Box_1 \Box_2}\over {\Box_3}} -
  {2{\Box_3}}
+ {{{{\Box_3}^3}}\over {\Box_1 \Box_2}}\Big)
R_1 R_2 R_3\hat{1}
\nonumber\\&&
+\frac1{18900}\Big({{{3{\Box_3}^3}}\over {2\Box_1\Box_2}}
-{{{2{\Box_1}^2}}\over {\Box_3}}
- {{\Box_1 \Box_2}\over {\Box_3}} -
  {3{\Box_3}\over {2}}
\Big)
R_{1\,\alpha}^\mu R_{2\,\beta}^{\alpha} R_{3\,\mu}^\beta\hat{1}
+\frac1{151200}\Big(\Box_1
- {{{{\Box_1}^2}}\over {\Box_2}}
\nonumber\\&&
+6\Box_3 + {8{\Box_1 \Box_3}\over {\Box_2}} -
  {{13 {{\Box_3}^2}}\over { \Box_2}}
+ {{{3{\Box_3}^3}}\over{\Box_1 \Box_2}}\Big)
R_1^{\mu\nu}R_{2\,\mu\nu}R_3\hat{1}
+{1\over {252}}
\hat{\cal R}_1^{\alpha\beta}\nabla^\mu\hat{\cal R}_{2\mu\alpha}
\nabla^\nu\hat{\cal R}_{3\nu\beta}
\nonumber\\&&
+{1\over {60}}
 \hat{\cal R}_1^{\mu\nu} \nabla_\mu\hat{P}_2\nabla_\nu\hat{P}_3
+{1\over {180}}
 \nabla_\mu \hat{\cal R}_1^{\mu\alpha}\nabla^\nu
\hat{\cal R}_{2\,\nu\alpha}\hat{P}_3
-{1\over {1890}}
 R_1^{\mu\nu}\nabla_\mu R_2\nabla_\nu \hat{P}_3
\nonumber\\&&
+\frac1{420}\Big({2\over3} + {{\Box_1}\over {\Box_2}}
+ {2{\Box_3}\over {\Box_2}} -
  {{3{{\Box_3}^2}}\over {2\Box_1 \Box_2}}\Big)
\nabla^\mu R_1^{\nu\alpha}\nabla_\nu R_{2\,\mu\alpha}\hat{P}_3
+{1\over {180}}
R_1^{\mu\nu}\nabla_\mu\nabla_\nu\hat{P}_2\hat{P}_3
\nonumber\\&&
+\frac1{105}\Big({1\over {12}} + {{\Box_3}\over {\Box_1}}\Big)
R_{1\,\alpha\beta}\nabla_\mu\hat{\cal R}_2^{\mu\alpha}
\nabla_\nu\hat{\cal R}_3^{\nu\beta}
+\frac1{210}\Big(-{1\over 6} - {{\Box_3}\over{\Box_1}}\Big)
R_1^{\alpha\beta}\nabla_\alpha\hat{\cal R}_2^{\mu\nu}
\nabla_\beta\hat{\cal R}_{3\,\mu\nu}
\nonumber\\&&
-{1\over {7560}}
R_1\nabla_\alpha\hat{\cal R}_2^{\alpha\mu}
\nabla^\beta\hat{\cal R}_{3\,\beta\mu}
+\frac1{105}\Big({1\over {6}} + {\Box_2\over {\Box_1}}
+ {\Box_3\over {\Box_1}}\Big)
R_1^{\mu\nu}\nabla_\mu\nabla_\lambda
\hat{\cal R}_2^{\lambda\alpha}\hat{\cal R}_{3\,\alpha\nu}
\nonumber\\&&
+\frac1{25200}\Big({1\over9} - {3{\Box_1}\over{\Box_2}}
-  {{{5{\Box_1}^2}}\over {2 \Box_2 \Box_3}}
+ {{\Box_3}\over { \Box_1}} -
  {{\Box_3}\over { \Box_2}}
- {{{{\Box_3}^2}}\over { \Box_1 \Box_2}}\Big)
R_1^{\alpha\beta}\nabla_\alpha R_2 \nabla_\beta R_3\hat{1}
\nonumber\\&&
+\frac1{37800}\Big({7{\Box_3}
\over {\Box_2}}-{{\Box_1}\over {\Box_2}}  -
  {{{3{\Box_3}^2}}\over {\Box_1 \Box_2}}\Big)
\nabla^\mu R_1^{\nu\alpha}\nabla_\nu R_{2\,\mu\alpha}R_3\hat{1}
\nonumber\\&&
+\frac1{12600}\Big(-{4\over3} - {{\Box_1}\over {\Box_2}} +
  {{{3{\Box_1}^2}}\over{2 \Box_2 \Box_3}}
- {2{\Box_3}\over{\Box_2}}\Big)
R_1^{\mu\nu}\nabla_\mu R_2^{\alpha\beta}\nabla_\nu
R_{3\,\alpha\beta}\hat{1}
\nonumber\\&&
+\frac1{9450}\Big(-3 - {{\Box_1}\over {\Box_2}} -
  {{{3{\Box_1}^2}}\over {2\Box_2 \Box_3}}
- {3{\Box_3}\over {\Box_1}} +
  {{\Box_3}\over {\Box_2}}
+ {{{3{\Box_3}^2}}\over{\Box_1 \Box_2}}\Big)
R_1^{\mu\nu}\nabla_\alpha R_{2\,\beta\mu}\nabla^\beta
R_{3\,\nu}^\alpha\hat{1}
\nonumber\\&&
+\frac1{420}\Big({1\over {\Box_2}}
+ {3{\Box_3}\over {2\Box_1 \Box_2}}\Big)
\nabla_\alpha\nabla_\beta R_1^{\mu\nu}\nabla_\mu
\nabla_\nu R_2^{\alpha\beta}\hat{P}_3
\nonumber\\&&
+\frac1{25200}\Big(-{{20}\over {3\Box_2}} - {4\over{\Box_3}} -
  {6{\Box_1}\over {\Box_2 \Box_3}}
+ {{\Box_3}\over {\Box_1 \Box_2}}\Big)
\nabla_\alpha\nabla_\beta R_1^{\mu\nu}\nabla_\mu
\nabla_\nu R_2^{\alpha\beta} R_3\hat{1}
\nonumber\\&&
+\frac1{6300}\Big(-{{4}\over{\Box_2}} - {8\over {3 \Box_3}}
-  {4{\Box_1}\over {\Box_2 \Box_3}}
- {{\Box_3}\over {\Box_1 \Box_2}}\Big)
\nabla_\mu R_1^{\alpha\lambda} \nabla_\nu
R_{2\,\lambda}^\beta\nabla_\alpha\nabla_\beta R_3^{\mu\nu}\hat{1}
\nonumber\\&&\left.
+{1\over {6300 \Box_1 \Box_2}}
 \nabla_\lambda\nabla_\sigma R_1^{\alpha\beta}
\nabla_\alpha\nabla_\beta R_2^{\mu\nu}\nabla_\mu
\nabla_\nu R_3^{\lambda\sigma}\hat{1} \right\}
+{\rm O}[\Re^4].                         \label{eqn:3.8}
\end{eqnarray}
\arraycolsep=3pt
\mathindent=\leftmargini

The task is now to bring expressions (\ref{eqn:3.6})--(\ref{eqn:3.8})
to a
local form
by restoring the Riemann tensor. The procedure that we use
is as follows. For each $a_n$, we first consider a linear combination
of all possible local invariants of the appropriate dimension with
unknown coefficients. Next, in this combination, we exclude the
Riemann tensor by the technique of Ref.\onlinecite{JMP} (see eq.(2.2)
of this
reference), and equate the result to the nonlocal expression
above. This gives a set of equations for the coefficients,
which, in
each case, has a {\em unique} solution. In the case of $a_2$,
there is
only one local invariant with explicit participation of the Riemann
tensor:
$\int\! dx\, g^{1/2}\,
R_{\alpha\beta\mu\nu}
R^{\alpha\beta\mu\nu}$.
In the case of $a_3$, there are seven (the integral over space-time
is assumed):
\mathindent = 0pt
\arraycolsep=0pt
\begin{equation}
\begin{array}{llll}
{\rm tr} {\hat{P} R^{\mu\nu\alpha\beta}
          R_{\mu\nu\alpha\beta}},
& \ \ \ \
  {\rm tr}  {\hat{\cal R}^{\alpha\beta}\hat{\cal R}^{\mu\nu}
               R_{\alpha\beta\mu\nu}},
& \ \ \ \
               {R^{\alpha\beta}_{\ \ \mu\nu}
                     R^{\mu\nu}_{\ \ \sigma\rho}
                         R^{\sigma\rho}_{\ \ \alpha\beta}},
& \ \ \ \
		{R^{\alpha\ \beta}_{\ \mu \ \nu}
                     R^{\mu \ \nu}_{\ \sigma \ \rho}
                         R^{\sigma \ \rho}_{\ \alpha \ \beta}},

\\
                {R^{\alpha}_{\beta}
                     R_{\alpha\mu\nu\sigma}
                         R^{\beta\mu\nu\sigma}},
& \ \ \ \
  {R R^{\mu\nu\alpha\beta}
          R_{\mu\nu\alpha\beta} },
& \ \ \ \
                {R^{\alpha\mu}
                    R^{\beta\nu}
                         R_{\alpha\beta\mu\nu}},

\end{array}                            \label{eqn:3.9}
\end{equation}
and the coefficient of the sixth turns out to be zero. In the
case of
$a_4$, there are ten (counting only cubic):
\begin{equation}
\begin{array}{llll}
{\rm tr}    {\Box\hat{P}
                 R^{\alpha\beta\mu\nu}
                            R_{\alpha\beta\mu\nu}},
& \ \
{\rm tr}    {\hat{P}
                 \nabla_\mu\nabla_\alpha R_{\nu\beta}
                         R^{\mu\nu\alpha\beta}},
& \ \
{\rm tr}    { \hat{\cal R}^{\alpha\beta}
                   \Box
                    \hat{\cal R}^{\mu\nu}
                           R_{\alpha\beta\mu\nu}},
& \ \
            {\Box R^{\alpha}_{\beta}
                         R_{\alpha\mu\nu\sigma}
                                 R^{\beta\mu\nu\sigma}},

\\
    {\Box R
        R^{\mu\nu\alpha\beta}
          R_{\mu\nu\alpha\beta}},
& \ \
               {R_{\mu\nu}
                       \nabla^\mu R_{\alpha\beta\sigma\rho}
                 \nabla^\nu R^{\alpha\beta\sigma\rho}},
& \ \
    {R
	\nabla_\mu\nabla_\alpha R_{\nu\beta}
                     R^{\mu\nu\alpha\beta}},
& \ \
    {\nabla_\mu R_{\nu\alpha}
		      \nabla^\alpha R_{\rho\sigma}
                            R^{\mu\rho\nu\sigma}},
\\
    { \nabla_\alpha R_{\beta\lambda}
		      \nabla_\mu R_{\nu}^{\lambda}
                                  R^{\alpha\beta\mu\nu}},
& \ \
    { R^{\alpha\mu}
	    \Box
		 R^{\beta\nu}
                      R_{\alpha\beta\mu\nu}},
\end{array}                                       \label{eqn:3.10}
\end{equation}
\arraycolsep=3pt
\mathindent=\leftmargini
and the last one proves to be absent. The number of invariants with
the Riemann tensor does not grow fast owing to the
Bianchi identities and, particularly, their corollary
which excludes $\Box  R_{\alpha\beta\mu\nu}$ in a local way.

The final results are as follows. The expressions (\ref{eqn:3.4}) and
(\ref{eqn:3.5}) are already in the local form. The expression
(\ref{eqn:3.6})
is brought to a local form
\begin{eqnarray}
\int\! dx\, g^{1/2}\,{\rm tr}\,\hat{a_2}(x,x)\!&=&\!
\int\! dx\,
g^{1/2}\,  {\rm tr}\left\{ \frac12 \hat{P}\hat{P}+
\frac1{12}\hat{\cal R}_{\mu\nu}
\hat{\cal R}^{\mu\nu}
\right.
\nonumber\\&&
\left.
+\,\left[\frac1{180} R_{\alpha\beta\mu\nu}
R^{\alpha\beta\mu\nu}
-\frac1{180} R_{\mu\nu}
R^{\mu\nu}\right]\hat{1}\,\right\}
+{\rm O}[\,\Re^4\,],                              \label{eqn:3.11}
\end{eqnarray}
by using the identity (6.36) of Ref.\onlinecite{JMP} which expresses
for
arbitrary spacetime dimension $2\omega$ the Gauss-Bonnet combination
of
Riemann and Ricci curvatures in terms of nonlocal invariants built of
the Ricci
tensor.

Finally, the expressions (\ref{eqn:3.7}), (\ref{eqn:3.8}) rewritten
in terms of
invariants
(\ref{eqn:3.9}) and (\ref{eqn:3.10}) take the form
\mathindent=0pt
\arraycolsep=0pt
\begin{eqnarray}&&
\int\! dx\, g^{1/2}\,  {\rm tr}\,\hat{a_3}(x,x)=\int\! dx\,
g^{1/2}\, {\rm tr} \left\{\frac{1}{12}\hat{P}\Box\hat{P}
+\frac{1}{120}\hat{\cal R}_{\mu\nu}\Box\hat{\cal R}^{\mu\nu}\right.
+\frac{1}{180}\hat{P}\Box R
\nonumber\\&&\ \ \ \ \ \ \ \
+\left[\frac{1}{840}R_{\mu\nu}\Box R^{\mu\nu}
-\frac{1}{3780}R\Box R\right]\hat{1}
%\nonumber\\&&\ \ \ \ \ \ \ \
+{1\over 6}\hat{P}\hat{P}\hat{P}
-{1\over {45}}
  	{\hat{\cal R}^{\mu}_{\ \alpha}
  		\hat{\cal R}^{\alpha}_{\ \beta}
\hat{\cal R}^{\beta}_{\ \mu}}
%\nonumber\\&&\ \ \ \ \ \ \ \
+{1\over {12}}
   {\hat{P}\hat{\cal R}^{\alpha\beta}\hat{\cal R}_{\alpha\beta}}
\nonumber\\&&\ \ \ \ \ \ \ \
+{1\over {72}}
    {R^{\mu\nu}_{\ \ \alpha\beta}
          \hat{\cal R}^{\alpha\beta}\hat{\cal R}_{\mu\nu}}
-{1\over {180}}
   {R^{\mu\nu}\hat{\cal R}^{\alpha}_{\ \mu}\hat{\cal R}_{\alpha\nu}}
%\nonumber\\&&\ \ \ \ \ \ \ \
+{1\over {180}}
    {R^{\mu\nu\alpha\beta}
          R_{\mu\nu\alpha\beta}\hat{P}}
%\nonumber\\&&\ \ \ \ \ \ \ \
-{1\over {180}}
    {R^{\alpha\beta}
          R_{\alpha\beta}\hat{P}}
\nonumber\\&&\ \ \ \ \ \ \ \
+\left[-\frac{1}{1620}
		{R^{\alpha\ \beta}_{\ \mu \ \nu}
                     R^{\mu \ \nu}_{\ \sigma \ \rho}
                         R^{\sigma \ \rho}_{\ \alpha \ \beta}}
\right.
%\nonumber\\&&\ \ \ \ \ \ \ \
+\frac{17}{45360}
               {R^{\alpha\beta}_{\ \ \mu\nu}
                     R^{\mu\nu}_{\ \ \sigma\rho}
                         R^{\sigma\rho}_{\ \ \alpha\beta}}
+\frac{1}{7560}
               {R_{\alpha\beta}
                     R_{\ \mu\nu\lambda}^{\alpha}
                         R^{\beta\mu\nu\lambda}}
\nonumber\\&&\ \ \ \ \ \ \ \ \left.\left.
+\frac{1}{945}
               {R_{\alpha\beta}
                     R^{\mu\nu}
                         R^{\alpha\ \beta}_{\ \mu \ \nu}}
-\frac{4}{2835}
               {R^{\alpha}_{\beta}
                     R^{\beta}_{\mu}
                         R^{\mu}_{\alpha}}
\right]\hat{1}\right\}+{\rm O}[\Re^4],             \label{eqn:3.12}
\end{eqnarray}
\begin{eqnarray}&&
\int\! dx\, g^{1/2}\,  {\rm tr}\,\hat{a_4}(x,x)=
\int\! dx\, g^{1/2}\, {\rm tr}
\left\{\frac1{120}\hat{P}{\Box^2}\hat{P}
+\frac1{1260}\hat{P}{\Box^2}R
\right.
+\frac1{1680}\hat{\cal R}^{\mu\nu}{\Box^2}\hat{\cal R}_{\mu\nu}
\nonumber\\&&\ \ \ \ \ \ \ \
+\frac1{15120}R^{\mu\nu}{\Box^2}R_{\mu\nu}\hat{1}
+\frac1{24}\Box\hat{P}\hat{P}\hat{P}
-\frac1{630}\Box
  	{\hat{\cal R}^{\mu}_{\ \alpha}
\hat{\cal R}^{\alpha}_{\ \beta}\hat{\cal R}^{\beta}_{\ \mu}}
+\frac1{252}
	{\hat{\cal R}^{\alpha\beta}
\nabla^\mu\hat{\cal R}_{\mu\alpha}\nabla^\nu\hat{\cal R}_{\nu\beta}}
\nonumber\\&&\ \ \ \ \ \ \ \
+\frac1{180}
{\Box\hat{\cal R}^{\mu\nu}\hat{\cal R}_{\mu\nu}\hat{P}}
+\frac1{180}
	{\hat{\cal R}^{\mu\nu}\Box\hat{\cal R}_{\mu\nu}\hat{P}}
+\frac1{90}
	{\hat{\cal R}^{\mu\nu}\hat{\cal R}_{\mu\nu}\Box\hat{P}}
%\nonumber\\&&\ \ \ \ \ \ \ \
+\frac1{180}
  	{\nabla_\mu \hat{\cal R}^{\mu\alpha}\nabla^\nu
\hat{\cal R}_{\nu\alpha}\hat{P}}
\nonumber\\&&\ \ \ \ \ \ \ \
+\frac1{720}
       \hat{P}\hat{P}\Box R
+\frac1{180}
	{R^{\mu\nu}\nabla_\mu\nabla_\nu\hat{P}\hat{P}}
+\frac1{60}
 	{\hat{\cal R}^{\mu\nu} \nabla_\mu\hat{P}\nabla_\nu\hat{P}}
-\frac1{1890}
	{R^{\mu\nu}\nabla_\mu R\nabla_\nu \hat{P}}
\nonumber\\&&\ \ \ \ \ \ \ \
-\frac1{15120}
      {\Box\hat{P} R R}
+\frac1{7560}
      {\hat{P} R\Box R}
-\frac1{1260}
 	{\nabla^\mu R^{\nu\alpha}\nabla_\nu R_{\mu\alpha}\hat{P}}
-\frac1{840}
	{R^{\mu\nu}\Box R_{\mu\nu}\hat{P}}
\nonumber\\&&\ \ \ \ \ \ \ \
-\frac1{5040}
	{R^{\mu\nu} R_{\mu\nu}\Box \hat{P}}
+\frac1{1120}
    {R^{\mu\nu\alpha\beta}
          R_{\mu\nu\alpha\beta}\Box\hat{P}}
+\frac1{420}
    {R^{\mu\nu\alpha\beta}
		 \nabla_\mu\nabla_\alpha R_{\nu\beta}
				\hat{P}}
\nonumber\\&&\ \ \ \ \ \ \ \
+\frac{13}{30240}
	{\Box R\hat{\cal R}^{\mu\nu}\hat{\cal R}_{\mu\nu}}
-\frac{1}{15120}
	{R\hat{\cal R}^{\mu\nu}\Box\hat{\cal R}_{\mu\nu}}
-\frac{1}{7560}
	{R\nabla_\alpha\hat{\cal R}^{\alpha\mu}
\nabla^\beta\hat{\cal R}_{\beta\mu}}
\nonumber\\&&\ \ \ \ \ \ \ \
+\frac{1}{840}
	{\Box R^{\alpha\beta}\hat{\cal R}_{\alpha}^{\ \mu}
        \hat{\cal R}_{\beta\mu}}
-\frac{1}{1260}
	{R^{\alpha\beta}\nabla_\alpha
\hat{\cal R}^{\mu\nu}\nabla_\beta
        \hat{\cal R}_{\mu\nu}}
+\frac{1}{630}
{R^{\mu\nu}\nabla_\mu\nabla_\lambda\hat{\cal R}^{\lambda\alpha}
     \hat{\cal R}_{\alpha\nu}}
\nonumber\\&&\ \ \ \ \ \ \ \
+\frac{1}{1260}
{R_{\alpha\beta}\nabla_\mu\hat{\cal R}^{\mu\alpha}\nabla_\nu
        \hat{\cal R}^{\nu\beta}}
+\frac{1}{420}
    {R_{\mu\nu\alpha\beta}
         \hat{\cal R}^{\alpha\beta}\Box\hat{\cal R}^{\mu\nu}}
\nonumber\\&&\ \ \ \ \ \ \ \
+\left[\,
\frac{1}{50400}
    {R^{\mu\nu\alpha\beta}
          R_{\mu\nu\alpha\beta}\Box R}
+\frac{1}{6300}
               {\Box R_{\alpha\beta}
                         R_{\ \mu\nu\lambda}^{\alpha}
                                 R^{\beta\mu\nu\lambda}}\!
-\frac{1}{25200}
               {R_{\lambda\sigma}
                       \nabla^\lambda R^{\mu\nu\alpha\beta}
                              \nabla^\sigma R_{\mu\nu\alpha\beta}}
\right.
\nonumber\\&&\ \ \ \ \ \ \ \
-\frac1{37800}
    {R^{\mu\nu\alpha\beta}
		 \nabla_\mu\nabla_\alpha R_{\nu\beta}
				R}
-\frac1{6300}
    {R^{\mu\alpha\nu\beta}
		 \nabla_\mu R_{\nu\lambda}
		      \nabla^\lambda R_{\alpha\beta}}
-\frac2{4725}
    {R^{\alpha\beta\mu\nu}
		 \nabla_\alpha R_{\beta\lambda}
		      \nabla_\mu R_{\nu}^{\lambda}}
\nonumber\\&&\ \ \ \ \ \ \ \
+\frac1{37800}
	{R^{\mu\nu}
               \nabla_\alpha R_{\beta\mu}
                       \nabla^\beta R_{\nu}^\alpha}
-\frac1{9450}
	{R^{\mu\nu}
                \nabla_\mu R^{\alpha\beta}
                          \nabla_\nu R_{\alpha\beta}}
-\frac1{18900}
	{\nabla^\mu R^{\nu\alpha}\nabla_\nu R_{\mu\alpha}R}
\nonumber\\&&\ \ \ \ \ \ \ \
+\frac{29}{453600}
	{R^{\alpha\beta}\nabla_\alpha R \nabla_\beta R}
+\frac{1}{37800}
	{R R^{\mu\nu}\Box R_{\mu\nu}}
-\frac{1}{75600}
	{\Box R R^{\mu\nu}R_{\mu\nu}}
\nonumber\\&&\ \ \ \ \ \ \ \
\left.\left.
-\frac{1}{7560}
  	{\Box R^{\mu}_{\alpha}
  		R^{\alpha}_{\beta} R^{\beta}_{\mu}}
-\frac{1}{100800}
	       {\Box R R R}
\right]\hat{1}\right\}+{\rm O}[\Re^4].           \label{eqn:3.13}
\end{eqnarray}
\arraycolsep=3pt
\mathindent=\leftmargini

The expressions (\ref{eqn:3.4}), (\ref{eqn:3.5}) and (\ref{eqn:3.11})
for $a_0,
a_1$ and $a_2$
coincide with the results obtained by other methods
\cite{2,5',7,8,9,10,3}.
It is easy to compare expression (\ref{eqn:3.12}) for $a_3$
with the result in \cite{10,17} since
they differ only by the substitution \cite{f11}
\begin{eqnarray}&&
\int\! dx\, g^{1/2}\,  {\rm tr}
\nabla_\alpha \hat{\cal R}^{\alpha\mu}
\nabla^\beta \hat{\cal R}_{\beta\mu}
=\int\! dx\, g^{1/2}\,  {\rm tr}\Big(
-\frac12 \hat{\cal R}_{\mu\nu}\Box \hat{\cal R}^{\mu\nu}
+2 {\hat{\cal R}^{\mu}_{\ \alpha}
   \hat{\cal R}^{\alpha}_{\ \beta}\hat{\cal R}^{\beta}_{\ \mu}}
\nonumber\\&& \qquad\qquad\qquad
 +    {R^{\mu\nu}\hat{\cal R}^{\alpha}_{\ \mu}
\hat{\cal R}_{\alpha\nu}}
- \frac12
    {R^{\mu\nu}_{\ \ \alpha\beta}
          \hat{\cal R}^{\alpha\beta}\hat{\cal R}_{\mu\nu}}
\Big),
\end{eqnarray}
and it is very difficult to compare expression (\ref{eqn:3.13}) for
$a_4$ with
the result in \cite{17}. An algebra of the Bianchi identities, Jacobi
identities for the commutator curvature (\ref{eqn:2.4}) and
integration by
parts, which
must be used for this purpose, can be found in
Refs.\onlinecite{JMP,16}. The
coincidence {\em
does} take place with
accuracy ${\rm O}[\Re^4]$ but expression (\ref{eqn:3.13}) is a result
of such
drastic simplifications  that it should be considered as {\em new}.
It goes without saying that, although all the equations
(\ref{eqn:3.11}) -- (\ref{eqn:3.13}) are presently obtained with
accuracy
${\rm O}[\Re^4]$, the results for $a_2$ and $a_3$ are exact. It is
also worth
emphasizing that the further expansion of the formfactors gives the
terms of
given quadratic and cubic orders in the curvature of {\it all} ${\rm
tr}\,\hat
a_n(x,x)$.

\section{The late-time behaviour of the trace of the heat kernel}
\indent
As discussed above, the late-time asymptotic behaviour is the most
important
result of covariant perturbation theory for the heat kernel, because
it gives a
universal criterion of the analyticity of the effective action
(\ref{eqn:1.5})
in the curvature for massless models and, thus, determines the range
of
applicability of this theory. Derivation of the late-time
behaviour of the form factors in the heat kernel was
given in paper \cite{14} to all orders in the curvature. For the
basic form factors (\ref{eqn:2.7}) and (\ref{eqn:2.8}) this behaviour
is
\begin{eqnarray}
&&f(-s\Box)=-\frac1s\frac2{\Box}
+{\rm O}\left(\frac1{s^2}\right),\ \ \ \
s\rightarrow\infty                      \label{eqn:4.1}\\
&&F(-s\Box_1,-s\Box_2,-s\Box_3)=
\frac1{s^2}\Big(\frac1{\Box_1\Box_2}
+\frac1{\Box_1\Box_3}+\frac1{\Box_2\Box_3}\Big)
+{\rm O}\left(\frac1{s^3}\right),
\ \ \ \  s\rightarrow\infty.        \label{eqn:4.2}
\end{eqnarray}
The late-time behaviours of all second-order and
third-order form factors follow then from expressions (\ref{eqn:4.1})
-
(\ref{eqn:4.2}) and the tables of paper \cite{16}.
One has
\arraycolsep=0pt
\mathindent=0pt
\begin{eqnarray}&&
f_i(-s\Box) =\frac{b_i}s\,\frac1{\Box} +
{\rm O}\left(\frac1{s^2}\right),\;\;
i=1\;{\rm to}\;5,\nonumber\\&&
\qquad
b_1=-1/6,\;b_2=1/18,\;b_3=1/3,
\;b_4=-1,\;b_5=-1/2,             \label{eqn:4.3}
\\&&
F_{i}(-s\Box_1,-s\Box_2,-s\Box_3)=\frac{a_i}{s^2}\left(\frac1{\Box_1
\Box_2}
+\frac1{\Box_1\Box_3}+\frac1{\Box_2\Box_3}\right)+{\rm
O}\left(\frac1{s^3}\right),\;
i=1\;{\rm to}\;7,9,10,\nonumber\\&&
\qquad
a_1=1/3,\;a_2=-2/3,\;a_3=0,\;a_4=1/36,\;a_5=0,
\;a_6=-1/6,
a_7=0,\;\nonumber\\&&\qquad
a_9=-1/648,\;a_{10}=0,           \label{eqn:4.4}
\\&&
F_{8}(-s\Box_1,-s\Box_2,-s\Box_3)
=\frac1{s^2}\left(\frac1{\Box_1\Box_2}
  +\frac1{\Box_1\Box_3}\right)+
  {\rm O}\left(\frac1{s^3}\right),         \label{eqn:4.5}
\\&&
F_{11}(-s\Box_1,-s\Box_2,-s\Box_3)
=\frac1{s^2}\frac1{12}\left(
  \frac1{\Box_1\Box_2}-\frac1{\Box_1\Box_3}
-\frac1{\Box_2\Box_3}\right)
  +{\rm O}\left(\frac1{s^3}\right),     \label{eqn:4.6}
  \\&&
 sF_{i}(-s\Box_1,-s\Box_2,-s\Box_3)
=\frac1{s^2}\frac{a_i}{\Box_1\Box_2\Box_3}
+{\rm O}\left(\frac1{s^3}\right),\;\;\;i=12\;{\rm to}\;25,
\nonumber\\&&\qquad
a_{12}=-2,\;a_{13}=-2,\;a_{14}=-2,\;a_{15}=0,\;a_{16}=0,\;a_{17}=0,\;
a_{18}=2,\;a_{19}=-1,\;
\nonumber\\&&\qquad
a_{20}=1/3,\;a_{21}=4,\;a_{22}=1/6,\;
a_{23}=-1/3,\;a_{24}=-1/3,\;a_{25}=0,    \label{eqn:4.7}
\\&&
  s^2F_{i}(-s\Box_1,-s\Box_2,-s\Box_3)
={\rm O}\left(\frac1{s^3}\right),\;\;
i=26,27,28,29.                            \label{eqn:4.8}
\end{eqnarray}
\arraycolsep=3pt
\mathindent=\leftmargini

The result is that the behaviour of the trace of the
heat kernel at large $s$ is $s^{-\omega+1}$, and the
coefficient of this asymptotic behaviour is obtained
to third order in the curvature. As shown in Ref.\onlinecite{14},
this
behaviour holds at all orders in the curvature except zeroth.
This power asymptotic behaviour is characteristic of
a non-compact manifold. For a compact manifold it will
be replaced by the exponential behaviour
${\rm Tr} K(s) \propto \exp(-\lambda_{\rm min}s),\,
s\rightarrow\infty$,
where $\lambda_{\rm min}$ is the minimum eigenvalue of the
operator $(-H)$ in (1.1). By applying the modification of
covariant perturbation theory, appropriate for
compact manifolds, one should be able to obtain this minimum
eigenvalue as a nonlocal expansion in powers of the
curvature (or a deviation of the curvature from the
reference one)\cite{f7}.

As seen from the expressions above, not all basis structures
contribute to the leading asymptotic behaviour. The
asymptotic form of ${\rm Tr} K(s)$ is as follows:
\mathindent=0pt
\arraycolsep=0pt
\begin{eqnarray}
{\rm Tr} K(s)&=&
{\frac{s}{(4\pi s)^\omega}\int\! dx\, g^{1/2}\,{\rm tr}\ }
\Big\{\hat{P}\!
%\nonumber\\&&
-\hat{P}\frac1{\Box}\hat{P}
	-\frac12\hat{\cal R}_{\mu\nu}
\frac1{\Box}\hat{\cal R}^{\mu\nu}\!
	+\frac13\hat{P}\frac1{\Box} R
	-\frac16R_{\mu\nu}\frac1{\Box} R^{\mu\nu}\hat{1}
	+\frac1{18}R\frac1{\Box} R\hat{1}
\nonumber\\&&
       +{\hat{P}\frac1{\Box}\hat{P}\frac1{\Box}\hat{P}}
       -2{\hat{\cal R}^{\mu}_{\ \alpha}\frac1{\Box}
\hat{\cal R}^{\alpha}_{\ \beta}
	\frac1{\Box}\hat{\cal R}^{\beta}_{\ \mu}}
%\nonumber\\&&
       +\frac1{36}{\frac1{\Box} R\frac1{\Box} R \hat{P}}
       +\frac1{18}{R\frac1{\Box} R\frac1{\Box} \hat{P}}
\nonumber\\&&
       -\frac16{\frac1{\Box}\hat{P}\frac1{\Box} \hat{P} R}
       -\frac13{\hat{P}\frac1{\Box} \hat{P}\frac1{\Box} R}
%\nonumber\\&&
       +2{\frac1{\Box}
R^{\alpha\beta}\frac1{\Box}\hat{\cal R}_{\alpha}^{\ \mu}
	\hat{\cal R}_{\beta\mu}}
       -\frac1{216}{R\frac1{\Box} R\frac1{\Box} R\hat{1}}
\nonumber\\&&
       +\frac1{12}{\frac1{\Box}
R^{\mu\nu}\frac1{\Box} R_{\mu\nu}R\hat{1}}
       -\frac16{R^{\mu\nu}\frac1{\Box}
R_{\mu\nu}\frac1{\Box} R\hat{1}}
       -2{\frac1{\Box}
\hat{\cal R}^{\alpha\beta}\nabla^\mu\frac1{\Box}
	\hat{\cal R}_{\mu\alpha}\nabla^\nu
\frac1{\Box}\hat{\cal R}_{\nu\beta}}
\nonumber\\&&
       -2{\frac1{\Box}\hat{\cal R}^{\mu\nu}
\nabla_\mu\frac1{\Box}\hat{P}\nabla_\nu\frac1{\Box}\hat{P}}
       -2{\nabla_\mu\frac1{\Box}\hat{\cal R}^{\mu\alpha}
	\nabla^\nu\frac1{\Box}
\hat{\cal R}_{\nu\alpha}\frac1{\Box}\hat{P}}
\nonumber\\&&
       +2{\frac1{\Box} R_{\alpha\beta}\nabla_\mu
	\frac1{\Box}\hat{\cal R}^{\mu\alpha}
\nabla_\nu\frac1{\Box}\hat{\cal R}^{\nu\beta}}
       -{\frac1{\Box} R^{\alpha\beta}\nabla_\alpha
	\frac1{\Box}\hat{\cal R}^{\mu\nu}\nabla_\beta
\frac1{\Box}\hat{\cal R}_{\mu\nu}}
\nonumber\\&&
       +\frac13{\frac1{\Box} R\nabla_\alpha
\frac1{\Box}\hat{\cal R}^{\alpha\mu}
	\nabla^\beta\frac1{\Box}\hat{\cal R}_{\beta\mu}}
       +4{\frac1{\Box} R^{\mu\nu}\nabla_\mu\nabla_\lambda
	\frac1{\Box}\hat{\cal R}^{\lambda\alpha}
\frac1{\Box}\hat{\cal R}_{\alpha\nu}}
\nonumber\\&&
       +\frac16{\frac1{\Box} R^{\alpha\beta}
	\nabla_\alpha\frac1{\Box} R \nabla_\beta
\frac1{\Box} R\hat{1}}
       -\frac13{\nabla^\mu\frac1{\Box} R^{\nu\alpha}
\nabla_\nu\frac1{\Box} R_{\mu\alpha}\frac1{\Box} R\hat{1}}
\nonumber\\&&
       -\frac13{\frac1{\Box} R^{\mu\nu}\nabla_\mu\frac1{\Box}
R^{\alpha\beta}\nabla_\nu\frac1{\Box} R_{\alpha\beta}\hat{1}}
%\nonumber\\&& \ \ \ \ \ \ \ \
 +{\rm O}[\Re^4]\Big\} +
{\rm O}\left(\frac1{s^\omega}\right),\ \ \ \
	s\rightarrow\infty.                 \label{eqn:4.9}
\end{eqnarray}
\arraycolsep=3pt
\mathindent=\leftmargini

As discussed in Introduction, the behaviour (\ref{eqn:4.9}) is very
important
for the effective action in massless theories. It controls the
convergence
of the integral (1.5) at the upper limit and serves as a criterion of
analyticity of the effective action in the curvature. For manifolds
of
dimension
\[2\omega>2\]
this integral converges at the upper limit at each order in the
curvature. Therefore, the effective action in four (and higher)
dimensions is
always analytic in the
curvature whereas in two dimensions, $\omega=1$, it is generally not.
An exceptional case is a conformal invariant two-dimensional scalar
field with
	\begin{equation}
	{\rm tr}\hat 1=1,\; \hat \Re_{\mu\nu}=0,\;
	\hat P=R/6,\;R_{\mu\nu}=g_{\mu\nu}R/2,\;
	g^{1/2}R={\rm a\;total\;derivative},\;
	\omega=1,                                 \label{eqn:4.15}
	\end{equation}
for which the effective action is expandable in
powers of the curvature because the integral (\ref{eqn:1.5})
converges at the
upper limit at each order of this expansion owing to specific
cancellations in
the leading asymptotic behaviour (\ref{eqn:4.9}) \cite{14,16}.
However, generally, for a two-dimensional theory the effective action
is
non-analytic in curvature, which implies that its calculation
requires a further summation of the curvature series in the heat
kernel -- a
technique which is beyond the presently discussed calculational
scheme (see
Introduction and Ref.\onlinecite{25}).

For a two-dimensional theory (\ref{eqn:4.15}) the full set of
curvature
invariants of eq.(\ref{eqn:2.6}) reduces to the following two
structures: $R_1
R_2$ and $R_1 R_2 R_3$, and ${\rm Tr} K(s)$ takes the form of an
expansion in
powers  of the  Ricci scalar only:
	\begin{eqnarray}
	{\rm Tr} K(s)\!&=&\!
	\frac1{4\pi s}\int\! dx\, g^{1/2}\,
	\left\{1+s^2\sum^{5}_{i=1}
	c_{i} f_{i}(-s\Box_2) R_1 R_2
	\right.\nonumber\\&& \qquad
 	\left.+s^3\sum^{29}_{i=1} C_{i}
	F_{i}(-s\Box_1,-s\Box_2,-s\Box_3)
	R_1 R_2 R_3 +{\rm O}[R^4]
 	\right\}, \ \ \ \  \omega=1            \label{eqn:4.16}
	\end{eqnarray}
where
\mathindent=0pt
\arraycolsep=0pt
\begin{eqnarray}
&&c_{1}=1/2,\ \ \ \  c_{2}=1,\ \ \ \  c_{3}=1/6,\ \ \ \
c_{4}=1/36,\ \ \ \  c_{5}=0,\nonumber\\
&&C_{1}=1/216,\ \
C_{4}=1/6,\ \
C_{5}=1/12,\ \
C_{6}=1/36,\ \
C_{9}=1,\ \
C_{10}=1/4,\ \
C_{11}=1/2,
\nonumber\\[1.5mm]&&
C_{15}=\frac{s}{24}(\Box_1-\Box_2-\Box_3),\ \
C_{16}=\frac{s}{48}(\Box_3-\Box_2-\Box_1),\ \
C_{17}=\frac{s}{72}\Box_2,
\nonumber\\[1.5mm]&&
C_{22}=\frac{s}{4}(\Box_1-\Box_2-\Box_3),\ \
C_{23}=\frac{s}{8}(\Box_3-\Box_2-\Box_1),\ \
C_{24}=\frac{s}{8}(\Box_1-\Box_2-\Box_3),
\nonumber\\[1.5mm]&&
C_{25}=\frac{s}{16}(\Box_1-\Box_2-\Box_3),\ \
C_{26}=\frac{s^2}{24}\Box_1\Box_2,\ \
 C_{27}=\frac{s^2}{4}\Box_1\Box_2,
\nonumber\\[1.5mm]&&
C_{28}=\frac{s^2}{16}\Box_3(\Box_3-\Box_2-\Box_1),\ \
C_{29}=\frac{s^3}{8}\Box_1\Box_2\Box_3,
\nonumber\\[1.5mm]&&
C_{2}=C_{3}=C_{7}=C_{8}=C_{12}=C_{13}=
C_{14}=C_{18}=C_{19}=C_{20}=C_{21}=0.            \label{eqn:4.17}
\end{eqnarray}
\arraycolsep=3pt
\mathindent=\leftmargini

By using in (\ref{eqn:4.16}) the asymptotic behaviours
(\ref{eqn:4.3}) --
(\ref{eqn:4.8}), one can now check that, at $s\rightarrow\infty$, the
leading
terms $1/s$ in ${\rm Tr} K(s)/s$ cancel at both second order and
third order in
the curvature so that
\begin{eqnarray}
\frac1s{\rm Tr} K(s)&=&{\rm O}\left(\frac1{s^2}\right),
\ \ \ \   s\rightarrow\infty.                    \label{eqn:4.18}
\end{eqnarray}
As a result, the integral (\ref{eqn:1.5}) converges at the upper
limit.
The convergence at the lower limit in the curvature-dependent terms
holds
trivially. Only the term of zeroth order in the curvature is
ultraviolet
divergent but, in the effective action of a massless theory, this
term gets
subtracted by a contribution of a local functional measure
\cite{Bern}.
The actual calculation of this integral can be performed by using in
(\ref{eqn:4.16}) the table of form factors of Ref.\onlinecite{16} and
the
differential equations (\ref{eqn:2.11}) -- (\ref{eqn:2.13}) for basic
form factors, which allow one to convert ${\rm Tr} K(s)/s$ into a
total
derivative in $s$:
\begin{eqnarray}
\frac1s{\rm Tr} K(s)\!\!&=&\!\!
\frac1{4\pi}  \frac{d}{ds}\int\! dx\, g^{1/2}\,\Big\{\!
-\frac1s+l\,(s,\Box_2) R_1 R_2
%\right.
\nonumber\\&&\qquad\qquad\qquad
%\left.
+h\,(s,\Box_1,\Box_2,\Box_3) R_1 R_2 R_3
+{\rm O}[R^4]\,\Big\},
\ \ \ \  \omega=1,                          \label{eqn:4.19}
\end{eqnarray}
where
\mathindent=0pt
\arraycolsep=0pt
\begin{eqnarray}&&
l\,(s,\Box)=\frac{1}{\Box}
\left[\,\frac18f(-s\Box)-\frac14
\frac{f(-s\Box)-1}{s\Box}\,\right],          \label{eqn:4.20}\\
&&
h(s,\Box_1,\Box_2,\Box_3)=
s F(-s\Box_1,-s\Box_2,-s\Box_3)
{{ {\Box_1} {\Box_2} {\Box_3}}\over{3 {{D}^2}}}+
\frac{f(-s\Box_1)}{8 {{D}^2}{\Box_1} {\Box_2}}\,
\Big ({{\Box_1}^4} \!- 2 {{\Box_1}^3} {\Box_3}
  \nonumber\\&&\ \ \ \
+2 {\Box_1} {{\Box_3}^3}\!- {{\Box_3}^4}\!- 2 {{\Box_1}^3} {{\Box_2}}
+3{\Box_2}{{\Box_3}^3}\!
- 8 {{\Box_1}^2}  {\Box_2} {\Box_3}
+8 {\Box_1} {\Box_2} {{\Box_3}^2}\!
- 10 {\Box_1} {{\Box_2}^2} {\Box_3}\! -2 {{\Box_2}^2}
{{\Box_3}^2}\Big )
\nonumber\\&&\ \ \ \
-\left[\,\frac{f(-s\Box_1)-1}{s{\Box_1}}\,\right]
\,\frac{1}{4 D{\Box_1} {\Box_2}}\,\Big({{\Box_1}^2} + 4 {\Box_1}
{\Box_2}
+ {\Box_2} {\Box_3} - {{\Box_3}^2}\Big)
\nonumber\\&&\ \ \ \
-\frac{1}{{\Box_2}\!-{\Box_3}}\frac{\Box_2}{\Box_1}\left\{
\frac18\left[\,\frac{f(-s\Box_2)}{\Box_2}
-\frac{f(-s\Box_3)}{\Box_3}\,\right]
-\frac14 \left[\,\frac{f(-s\Box_2)-1}{s{\Box_2}^2}
-\frac{f(-s\Box_3)-1}{s{\Box_3}^2}\,\right]\right\}.
\label{eqn:4.21}
\end{eqnarray}
\arraycolsep=3pt
\mathindent=\leftmargini

Insertion of (\ref{eqn:4.19}) (with the subtracted term of zeroth
order in $R$)
in (\ref{eqn:1.5}) gives for the effective action:
\begin{eqnarray}
W &=&
\frac1{8\pi}\int\! dx\, g^{1/2}\,
\Big\{\,l\,(0,\Box_2)R_1 R_2
%\nonumber\\&&
+h(0,\Box_1,\Box_2,\Box_3) R_1 R_2 R_3
+{\rm O}[R^4]\,\Big\},
\;\omega=1,                 \label{eqn:4.22}
\end{eqnarray}
where use is made of the fact that the functions
$l$ and $h$ vanish at
\hbox{$s\rightarrow\infty$}. With the asymptotic behaviours
(\ref{eqn:3.1}) and (\ref{eqn:3.2}), and the explicit expressions
above, we
obtain for $l\,(0,\Box)$ and the completely symmetrized in
$\Box_1,\Box_2,\Box_3$ function $h(0,\Box_1,\Box_2,\Box_3)$ (which
only
contribute to (\ref{eqn:4.22}))
\begin{eqnarray}
l\,(0,\Box)=\frac{1}{12}\frac{1}{\Box}, &&
h^{\rm sym}(0,\Box_1,\Box_2,\Box_3)=0,
\end{eqnarray}
whence
\begin{equation}
W=\frac1{96\pi}\int\! dx\, g^{1/2}\,
R\frac{1}{\Box}R+{\rm O}[R^4],\ \ \ \   \omega=1.
\end{equation}
Here the term of second order in the
curvature reproduces the result of paper \cite{14} (and the results
of
Refs.\onlinecite{6,Fr-Vil}
obtained by integrating the trace
anomaly).

Thus the third--order contribution in $W$
really vanishes, and the mechanism
of this vanishing is that, under special
conditions like (\ref{eqn:4.15}),
the third-order contribution in $s^{-1}{\rm Tr} K(s)$
becomes a total derivative
of a function vanishing at both $s=0$ and $s=\infty$.
This mechanism underlies
all "miraculous" cancellations of nonlocal
terms including the trace anomaly
in four dimensions.

\section*{Acknowledgments}

\indent
The work of A.O.B. on this paper was partially supported by the CITA
National Fellowship and NSERC grant at the University of Alberta.
G.A.V. has
been supported by the Russian Science Foundation under Grant
93-02-15594 and by
a NATO travel grant (CRG 920991). V.V.Z., whose work was
supported in part by National Science Council of the Republic of
China under contract No. NSC 82-0208-M008-070, thanks the Department
of Physics
of National Central University for the support in computing
facilities.


\begin{thebibliography}{99}

\bibitem{Fock}
V.Fock, Phys.Zeit.Soviet Un. {\bf 12}, 404 (1937)

\bibitem{1}
J. S. Schwinger, Phys. Rev. {\bf 82}, 664 (1951)

\bibitem{2}
B. S. DeWitt,
{\it Dynamical theory of groups and fields }
(Gordon and Breach, New York, 1965)

\bibitem{5'}
G. 't Hooft,
Nucl. Phys. {\bf B62}, 444 (1973);
G. 't Hooft and M. Veltman,
Ann. Inst. Henri Poincar\'e
{\bf XX}, 69 (1974)

\bibitem{18}
G. A. Vilkovisky, in Quantum Theory of Gravity,
ed. S.M.Christensen
(Hilger, Bristol, 1984)

\bibitem{3}
A. O. Barvinsky and G. A. Vilkovisky,
Phys. Reports {\bf 119}, 1 (1985)

\bibitem{4}
A. O. Barvinsky and G. A. Vilkovisky,
{\it Quantum field theory and quantum statistics},
vol. 1, eds. I. A. Batalin, C. J. Isham and
G. A. Vilkovisky
(Hilger, Bristol, 1987) p.245

\bibitem{5}
G. A. Vilkovisky,
Preprint CERN-TH.6392/92;
Publication de l'Institut de Recherche
Math\'ematique Avanc\'ee, R.C.P. 25, vol. 43
(Strasbourg, 1992) p.203

\bibitem{6}
A. M. Polyakov, Phys. Lett. {\bf B103}, 207 (1981)

\bibitem{7}
J. Hadamard,
{\it Lectures on Cauchy Problem}, in:
Linear partial differential equations
(Yale U.P., New Haven, CT, 1923)

\bibitem{8}
S. Minakshisundaram and A. Pleijel,
Can. J. Math. {\bf 1}, 242 (1949)

\bibitem{9}
H. P. McKean and I. M. Singer,
J. Diff. Geom. {\bf 1}, 43 (1967);
R. T. Seely,
Proc. Symp. Pure Math. {\bf 10}, 288 (1967);
M. Atiyh, R. Bott and V. K. Patodi,
Invent. Math. {\bf 19}, 279 (1973)

\bibitem{10}
P. B. Gilkey, J. Diff. Geom. {\bf 10}, 601 (1975)

\bibitem{11}
P. B. Gilkey,
{\it Invariance theory, the heat equation and the
Atiyh-Singer index theorem }
(Publish or Perish, Wilmington, DE, USA, 1984)

\bibitem{21}
Y.Fujiwara, T.A.Osborn and S.F.J.Wilk,
Phys.Rev. {\bf A25}, 14 (1982);
F.H.Molzahn and T.A.Osborn,
J.Phys. {\bf A20}, 3073 (1987); F.H.Molzahn, T.A.Osborn and
S.A.Fulling,
Ann.Phys. (N.Y.) {\bf 204}, 64 (1990)

\bibitem{12}
R. Camporesi, Phys. Reports {\bf 196}, 1 (1990)

\bibitem{24}
T.Branson, S.Fulling and P.B.Gilkey, J.Math.Phys. {\bf 32} (1991)
2089;
V.P.Gusynin, E.V.Gorbar  and V.V.Romankov, Nucl.Phys. {\bf B362}
(1991)
449; S.Fulling, {\it Kernel asymptotics of exotic second-order
operators},
Proceedings of 3rd International Colloquium on Differential
Equations, Plovdiv,
Bulgaria, August 1992 (to appear)

\bibitem{Ram}{\it Forty more years of ramifications: spectral
asymptotics and
its applications}, eds. S.A.Fulling and F.J.Narcowich, (Texas A\&M
University,
College Station, 1991)

\bibitem{13}
A. O. Barvinsky and G. A. Vilkovisky,
Nucl. Phys. {\bf B282}, 163 (1987)

\bibitem{14}
A. O. Barvinsky and G. A. Vilkovisky,
Nucl. Phys. {\bf B333}, 471 (1990)

\bibitem{15}
A. O. Barvinsky and G. A. Vilkovisky,
Nucl. Phys. {\bf B333}, 512 (1990)

\bibitem{16}
A. O. Barvinsky, Yu. V. Gusev, G. A. Vilkovisky
and V. V. Zhytnikov,
Covariant Perturbation Theory (IV). Third Order in the Curvature,
Report of the
University of Manitoba (University of Manitoba, Winnipeg, 1993)

\bibitem{F}
One can also mention the method of pseudodifferential operators of
Refs.\onlinecite{24,Ram}, the connected-graph and gauge-invariant
derivative
expansions of Ref.\onlinecite{21}, but these methods cannot be
regarded as
completely independent and regular, because they either represent the
version
of the early-time expansion \cite{24} or are applicable to a limited
class of
problems in flat spacetime with scalar-field operators \cite{21}.


\bibitem{17}
I.G.Avramidi,
Teor.Mat.Fiz. {\bf 79}, 219 (1989);
Phys.Lett. {\bf B238}, 92 (1990);
Nucl.Phys. {\bf B355}, 712 (1991)

\bibitem{Frolov-Zeln}
See, e.g. V.P.Frolov and A.I.Zelnikov, Phys.Lett. {bf 115B}, 372
(1982); {\bf
123B}, 197 (1983); Phys.Rev. {\bf D29}, 1057 (1984)

\bibitem{Gibbons}
G.W.Gibbons, in {\it General Relativity. An Einstein Centenary
Survey}, eds.
S.W.Hawking and W.Israel (Cambridge University Press, Cambridge,
1979) p.639

\bibitem{Amsterdamski}
For particular models in flat spacetime this coefficient was obtained
in
S.F.J.Wilk, Y.Fujiwara and T.A.Osborn, Phys.Rev. {\bf A24}, 2187;
A.I.Wainstein, V.I.Zakharov, V.A.Novikov and M.A.Shifman,
Sov.J.Nucl.Phys. {\bf
39}, 77 (1984). For a scalar field operator in curved spacetime it
was found in
P.Amsterdamski, A.Berkin and D.O'Connor, Class.Quantum Grav. {\bf 6}
(1989) 1981

\bibitem{f2}
This summation was proposed in Ref.\onlinecite{18} and subsequently
actually
realized to second order in the curvature in I.G.Avramidi, Yad.Fiz.
{\bf 49},
1185 (1989); Phys.Lett. {\bf B236},
443 (1990). As
distinct
from this, covariant perturbation theory \cite{13,14,15,16} does not
sum any
infinite series; it gives the results of such a summation but is a
regular method.

\bibitem{Hawking}
S.W.Hawking, Commun.Math.Phys. {\bf 43}, 199 (1975)

\bibitem{Fr-Vil}
V.P.Frolov and G.A.Vilkovisky, Proc. Second Seminar on Quantum
Gravity (Moscow,
1981), eds. M.A.Markov and P.C.West (Plenum, London, 1983) p.267;
Phys.Lett.
{\bf 106 B}, 307 (1981)

\bibitem{Vil-Tex}
G.A.Vilkovisky, Quantum theory of the gravitational collapse, {\it
Lectures at
the University of Texas at Austin, manuscript} (1989)

\bibitem{CQG}
G.A.Vilkovisky, Class.Quantum Grav. {\bf 9}, 895 (1992)

\bibitem{Armen}
A.G.Mirzabekian and G.A.Vilkovisky, Phys.Lett. {\bf B317}, 517 (1993)

\bibitem{JMP}
A.O.Barvinsky, Yu.V.Gusev, G.A.Vilkovisky
and V.V.Zhytnikov, The basis of nonlocal curvature invariants in
quantum
gravity theory. (Third order.), in press

\bibitem{files}
These files are available from the authors.


\bibitem{20}
A.O.Barvinsky and Yu.V.Gusev,
Class. Quantum Grav. {\bf 9}, 383 (1992)

\bibitem{22}
A.O.Barvinsky and T.A.Osborn,
A Phase-space Technique for the
Perturbation Expansion of Schrodinger
Propagators, preprint of the University of
Manitoba MANIT-92-01
(Winnipeg, 1992)

\bibitem{23a}
A.I.Zelnikov, Phys. Lett. {\bf B273}, 471 (1991)

\bibitem{23b}
A.V.Leonidov and A. I. Zelnikov,
Phys. Lett. {\bf B276}, 122 (1992)

\bibitem{Jack-P}
I.Jack and L.Parker, Phys.Rev. {\bf D31},2439 (1985)

\bibitem{25}
I.G.Avramidy, Phys. Lett. {\bf B305}, 27 (1993). See also the earlier
work of
I.A.Batalin, S.G.Matinyan and G.K.Savvidi, Sov.J.Nucl.Phys. {\bf 26},
407
(1977); G.K.Savvidi, Phys.Lett. {\bf B71}, 133 (1977)

\bibitem{Avramidi} I.G.Avramidi, private communication.

\bibitem{f4}
We use the conventions
$R^\mu_{\,\cdot\, \alpha\nu\beta}=\partial_\nu
\Gamma^\mu_{\alpha\beta}-\cdots,\  R_{\alpha\beta}=
R^\mu_{\,\cdot\, \alpha\mu\beta},\
R=g^{\alpha\beta}R_{\alpha\beta}$.


\bibitem{f6}
To second order in the curvature, the result for
${\rm Tr} K(s)$ was first published in
A.O.Barvinsky and G.A.Vilkovisky,
Proceedings of the Fourth Seminar on
Quantum Gravity,
May 25-29, 1987, Moscow,
eds. M.A.Markov, V.A.Berezin and V.P.Frolov
(World Scientific, Singapore, 1988) p.217.

\bibitem{f7}
One of the authors (G.A.V.) is grateful to Gary Gibbons for
discussing this
point.

\bibitem{f11}
Which in eq.(8.21) of Ref.\onlinecite{14} figures with the
cubic terms
omitted.

\bibitem{Bern}
E.S.Fradkin and G.A.Vilkovisky, Lett.Nuovo Cim. {\bf 19}, 47 (1977);
Report of
the Institute for Theoretical Physics at Bern (1976)

\end{thebibliography}
\end{document}